# NAME: A Naming Mechanism for Delay/Disruption-Tolerant Network

Songyang Wang, Ruyi Ma

School of Computer Science and Engineering, Beihang University
Beijing 100191, China

*ABSTRACT*

*This paper presents the design and implementation of the naming mechanism (NAME), a resource discovery and service location approach for Delay/Disruption-Tolerant Network (DTN). First discuss the architecture of NAME mainly including Name Knowledge Base, Name Dissemination, Name Resolution and Name-based Routing. In the design and implementation of NAME, we introduce the simple name-specifiers to describe name, the name-tree for name storage and the efficient predicate-based routing algorithm. Future work is finally discussed for completing NAME and providing APIs for abundant applications.*

## 1 Introduction

Delay/Disruption-Tolerant Network (DTN) is an architecture and a set of protocols that enable communication in environments with intermittent connectivity and long delays [1]. Bundle Protocol (BP) [2] and an architecture for DTN [3] have been previously defined. To provide its services, BP sits at the application layer of some number of constituent internets, forming a store-and-forward overlay network. Key capabilities of BP include:

- Custody-based retransmission
- Ability to cope with intermittent connectivity
- Ability to take advantage of scheduled, predicted, and opportunistic connectivity (in addition to continuous connectivity)
- Late binding of overlay network endpoint identifiers to constituent internet addresses

In a DTN, nodes and services can appear, move, and disappear dynamically. That makes nodes impossible have the whole information about the state of addresses names and routes. After a bundle is created, the destination node may have changed. In such a flexible environment, it is significant to use name attributes of nodes (such as roles, location, or sensed values) and canonical DTN endpoint identifiers to locate nodes.

An Endpoint Identifier (EID) [3] is a name, expressed using the general syntax of Uniform Resource Identifier (URIs), that identifies a DTN endpoint. Using an EID, a node is able to determine the Minimum Reception Group (MRG) of the DTN endpoint named by the EID. The MRC of an endpoint may refer to one node (unicast) one of a group of nodes (anycast) or all of group of nodes (multicast and broadcast). Each node is also required to have at least on EID that uniquely identifies it. Canonical EID refers to the EID of a bundle processing entity that is capable of receiving bundles addressed to that EID from other DTN nodes. Every DTN node should possess a unique canonical EID. Naming mechanism's main target is to bind the name attributes to the canonical EID.

The important design goals of a naming system that enables resource discovery and service location are as follows:

- Expressiveness. The naming system must be flexible in order to handle a wide variety of devices and services. It must allow applications to express arbitrary service descriptions and queries.
- Responsiveness. The naming system must adapt quickly to end-node and service mobility, performance fluctuations, and other factors that can cause a change in the "best" network location of a service.
- Robustness. The naming system must be resilient to name resolver and service failures as well as inconsistencies in the internal state of the resolvers.
- Easy configuration. The name resolvers should configure themselves with minimal manual intervention and the system should not require manual registration of services. The resulting system should automatically distribute request resolution load among resolvers.

The main contribution of our work is the design and implementation of NAME, a naming mechanism for Delay/Disrupted-Tolerant network. The complete architecture of NAME, mainly including Name Knowledge Base, Name Dissemination, Name Resolution and Name-based Routing, is designed as a base for realizing NAME system. In the design and implementation of NAME, we introduce the simple name-specifiers to describe name, the name-tree for name storage and the efficient predicate-based routing algorithm. In future work, we will complete our DTN naming system, and design APIs for all kinds of applications such as content-based data dissemination and resource location.

The rest of this paper is organized as follows. Section 2 discusses the architecture of DTN naming system, Section 3, the design and implementation of naming mechanism NAME. Finally in Section 4, we conclude and introduce future work.

## 2 The Architecture of NAME

In this section, we discuss the whole architecture of NAME, which mainly consists of eight modules: Applications (APP), Naming system API, Database (DB), Name-based routing, Name resolution, Name knowledge base (NKB), Bundle protocol agent (BPA) and Convergence layer adaptor (CLA). The architecture is displayed in Fig.1.

### 2.1 Name Knowledge Base

Each node maintains a Name Knowledge Base for storing and processing ontology as well as attributes. NKB can contain information about bindings of name attributes to locally registered application End Identifiers (EIDs), or to remote related nodes' EIDs. The Name knowledge base can range from a simple table, matching name attributes to nodes' EIDs, to a powerful deductive database engine (such as Prolog) that can use predicate logic and negation as failure to infer complex derived attributes.

The range of possible knowledge base implementation is broad and is likely to be the subject of significant research and experimentation. NKB is made of the following parts [1]:

- A simple lookup table mapping ontology names (e.g. geographic) to nodes that have advertised the corresponding ontology. When an ontology advertisement arrives it is inserted into the table.

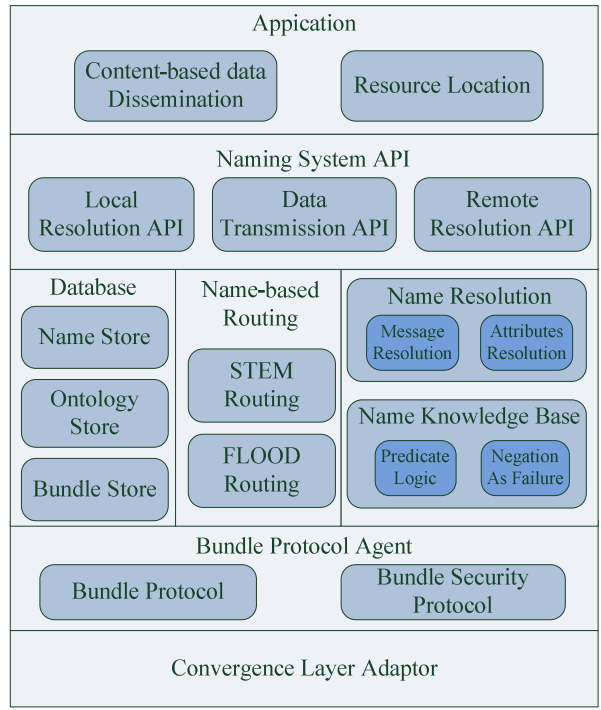

**Fig.1** The architecture of DTN naming system

- A lookup table matching name attributes to nodes' EIDs. The knowledge base can perform simple matches and comparisons to improve its forwarding (choice of related node) decisions. This mapping is possible if the corresponding ontology rules already reside on the current node.
- A deductive database that stores facts about name attributes as well as rules used for attributes derivation. This kind of KB can be used for table matching, as well as execute a potentially complex rule to infer the result from a given fact base.

### 2.2 Name Dissemination

The procedure of Name dissemination is as follows. Applications first register in the bundle protocol agent and add attribute' name and attribute's value by using name system API. The attributes' information is stored in the local name knowledge base. Based on the name information offered by applications, bundle protocol agent can infer and extent name information. The name knowledge base in every node should periodically disseminate information to other neighbor nodes.

The interaction among BPA, Resolver and Router is displayed in Fig.2.

1. An application registers in BPA
2. BPA sends name information to Resolver
3. Resolver updates the local NKB, extracts name attributes and resolves these attributes

4. Resolver sends name attributes and resolution information to Router
5. Router prepares to disseminate name based on the resolution information.

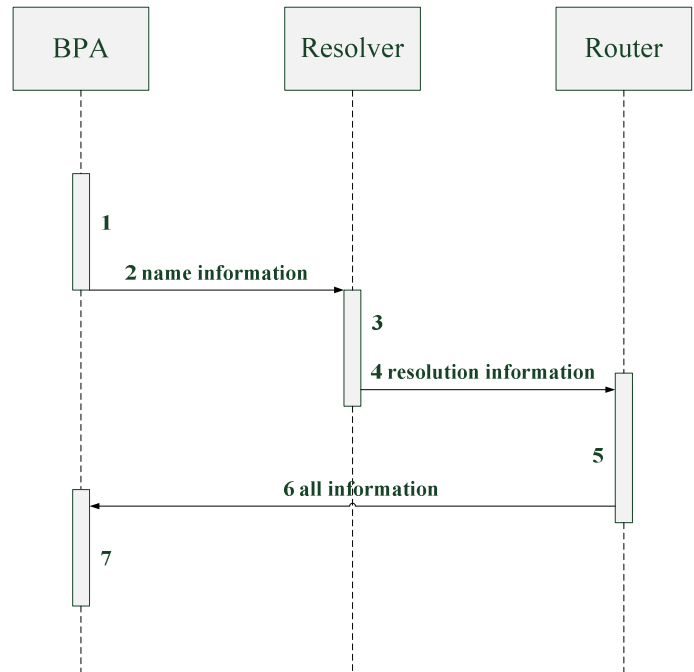

**Fig.2** the process of data dissemination

6. Router sends all the information to BPA
7. BPA creates a bundle and inserts it to CLA

### 2.3 Name Resolution

Name Resolution (also called as binding) is the process of matching a name to several destination nodes' EIDs, or to other names, or to related nodes' EIDs. In the process of name resolution, the bundle coming from local application or remote application is first stored in local DB, and then triggers the event BundleReceived. Router queries Resolver for the EIDs of the next related nodes and sends the bundle to these nodes. If some nodes cannot resolve the coming information, they should use bundle-in-bundle technique to transfer the bundle. In bundle-in-bundle technique, the nodes that are not able to resolve the current bundle could first encapsulate it in an "envelope bundle", then address this "envelope bundle" to the next resolver EID.

The interaction among BPA, Resolver and Router is displayed in Fig.3.

1. An application registers a bundle in BPA
2. Trigger the event BundleReceived
3. Resolver queries local name knowledge base and sends the information to Router
4. Router asks Resolver to resolve and infer name attributes
5. Resolver sends resolution information back to Router
6. Router sends all the information including routing approaches to BPA
7. BPA creates a bundle and inserts it to CLA

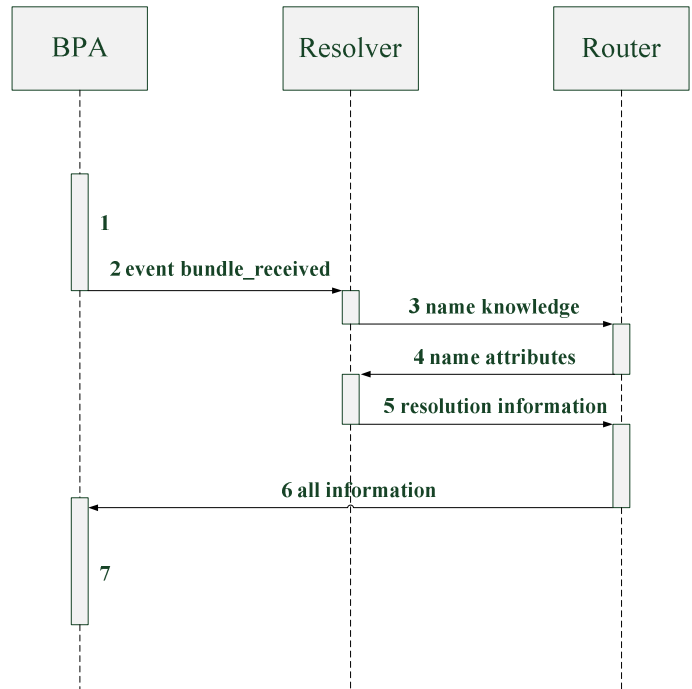

**Fig.3** The Process of Name Resolution

## 2.4 Name-based Routing

There have been numerous investigations about routing in a DTN. In a naming system, the main mission of DTN router is finding the destination nodes that match the name information in the bundle. DTN router should also make an appropriate decision to transfer a bundle based on the resolution clues. In all, the capability of DTN router directly determines the complex of a naming system.

Name is used to describe the features of destination nodes. Only by making full use of name information and other local information can router finish its function. Naming and routing are inseparable. The simplest routing is early-binding that determines the EIDs of destination nodes in the source node. The EIDs information is usually stored in the Metadata Extension Block (MEB). The whole process of name-based routing is in Fig.4:

When the application data reaches a DTN node, it first attempts to acquire the corresponding canonical EID of the destination name. If succeed, the node will bind the name to the canonical EID, and choose the next-hop node based on DTN routing decision. On the contrary, if the DTN node fails to acquire the canonical EID, it will narrow down the range of next-hop nodes by resolving name and reasoning name information, and then forward bundle based on DTN routing decision. The corresponding canonical EID can also be obtained through remote name server.

## 2.5 Bundle Protocol Agent (BPA)

BPA of a node is the node component that offers the BP services and executes the procedures of the bundle protocol. The manner in which it does so is wholly an implementation mater. For example, BPA functionality might be coded into each node individually; it might be implemented as a shared library that is used in common by any number of bundle nodes on a single computer; it might be implemented as a daemon whose services are invoked via inter-process or network communication by any number of bundle nodes on one or more computers; it might be implemented in hardware.

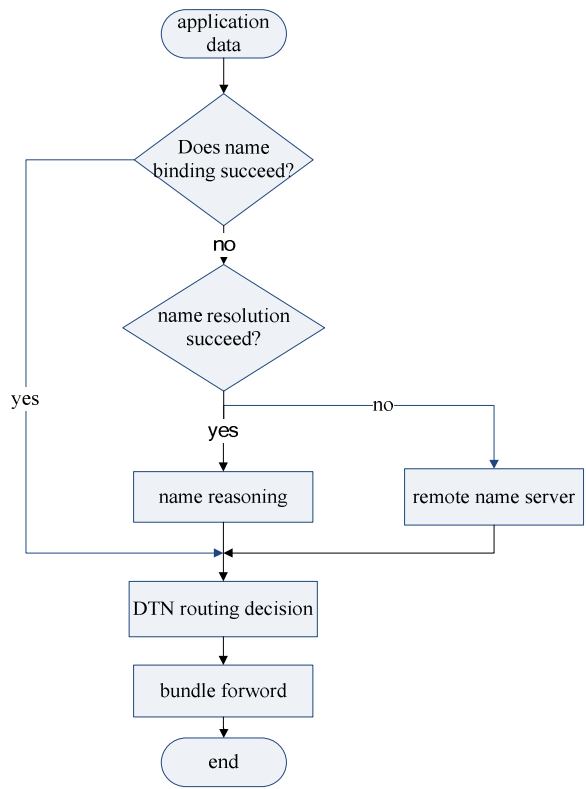

**Fig.4** The Process of Name-Based Routing

## 3    Design and Implementation of NAME

In this section, we present the details of design and implementation of our naming system based on the above architecture of DTN naming system.

### 3.1    Overview

Our naming mechanism (NAME) resolves name mainly based on the geographic information. We assume that every node has gotten their geographic information (longitude and latitude) through GPS device. The name in our system is made up of role attributes, location attributes and distance attributes, for example, "generals (role) located within 2 km (distance) of 116 degrees (longitude), 112 degrees (latitude)".

In the NAME, name resolution is completed during the transmission of a bundle instead of in the source node. At the beginning, NAME determines some closer nodes by using geographic information and transfer the bundle to them. This process will not stop until the bundle reaches the target region. When the bundle is within the target region, NAME will use other attributes to find the destination nodes.

The routing state can be STEM or FLOOD [4]. When it is STEM, NAME makes use of less network resource to send the bundle to the target region. When the bundle is within the target

236

International Journal of Computer Networks & Communications (IJCNC) Vol.5, No.6, November 2013

region, the routing state will change as FLOOD to take advantage of abundant resource. FLOOD is predicate-based epidemic routing algorithm. The process of routing is displayed in Fig.5.

In the predicate-based epidemic routing algorithm, we hope the bundle could reach the target region within (longitude, latitude, distance). This limitation can be realized by deploying predicate filter. During the implementation of NAME, the information of routing state and routing control will be stored in the MEB.

In NAME system, the MEB will have several formats as follows:

- Next resolver EID: This field denotes the next node that a bundle should be sent to. The routing algorithm is responsible for populating and computing this field.
- Routing state: Currently this state indicates whether the bundle should be sent "point-to-point" (if the Next resolver EID field is non-empty) or should the routing state be STEM or FLOOD (in this case, the Next resolver EID is likely to be empty).
- Conditions for scoping the transmission of the bundle: this field will contain a time-to-live value or predicate expressions that can govern the spatial scope of transmission.

### 3.2 Name Format

Our naming system uses name-specifiers [5] to express the meaning of a node's name in DTN. The source node puts name-specifies in the head of bundle to determine the desired destination node and source node. Name-specifiers are designed to be simple and easy to operate. The two main parts of the name-specifier are the attribute and the value. An attribute can classify an object into a category, for example, 'role'. A value is the object's classification within that category, for example, 'general'. Attributes and values are free-form strings that are defined by applications; name-specifiers do not restrict applications to using a fixed set of attributes and values. An attribute and its associated value together form an attribute-value pair or av-pair such as 'role = general'.

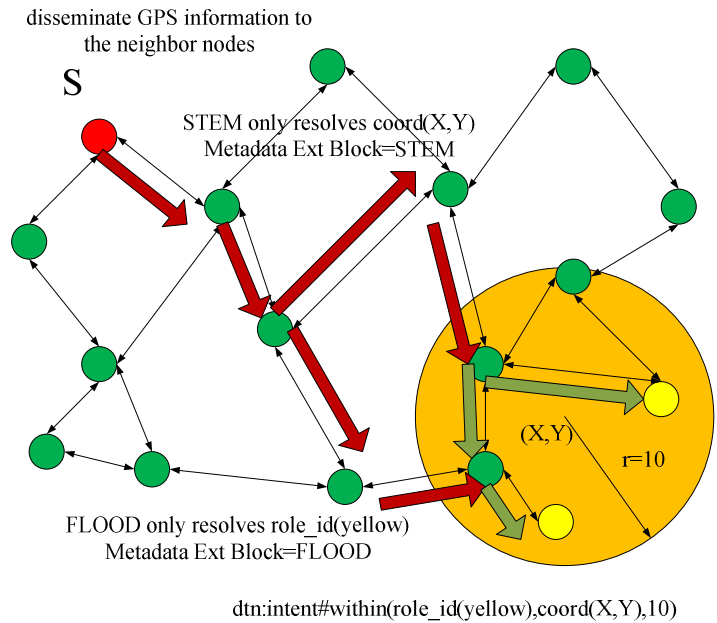

Fig.5 The process of routing in NAME

A name-specifier is a tree arrangement of av-pairs such that an av-pair only depends on its parent av-pairs. For instance, in the example name-specifier shown in Fig.6, a person's mission as command is meaningful only in the condition that this person's role is general, so the av-pair 'mission = command' depends on the 'role = general'; The siblings (av-pairs) in the tree are orthogonal to each other but dependent on the same parent av-pair. For example, a node's longitude and latitude can be selected independently of each other, and are meaningful only in the context of the location information. Therefore, the av-pairs 'longitude = 116 degrees' and 'latitude = 112 degrees' are independent. This tree arrangement makes name-specifiers easier to read and narrows down the search space during name resolution.

In the head of a bundle, the format of name-specifiers is shown in Fig.7 to describe source and destination of the bundle. This string-based representation is readable and helpful for debugging, in the spirit of other string-based protocols like HTTP [6] and NNTP [7]. The use of brackets '[' and ']' indicates levels of nesting, and av-pairs are separated by an equals sign '='. The arbitrary use of whites pace is permitted anywhere within the name specifier, except in the middle of attribute and value tokens.

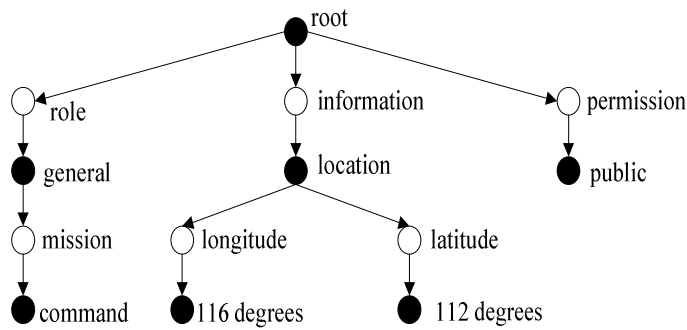

Fig.6 Name-specifier tree

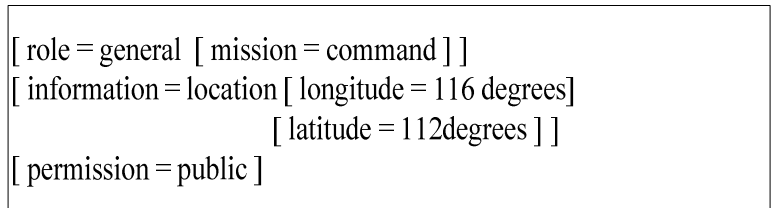

Fig.7 the format of name-specifier in a bundle

### 3.3 Name Storage

We use data structure Name-trees to store the correspondence between name-specifiers and name-records. The information of name-records contains the potential final destination node's EID, the next-hop nodes' EID and the life circle of name-records. Similar to the structure of name-specifers, Name-trees also consist of several av-pairs. But the difference is that the name-tree is a combination of all the name-specifiers a node knows about. So in the name-tree, an attribute can have many values. Each of these name-specifiers has a pointer from each of its leaf-values to a name-record. Fig.8 depicts an example name-tree.

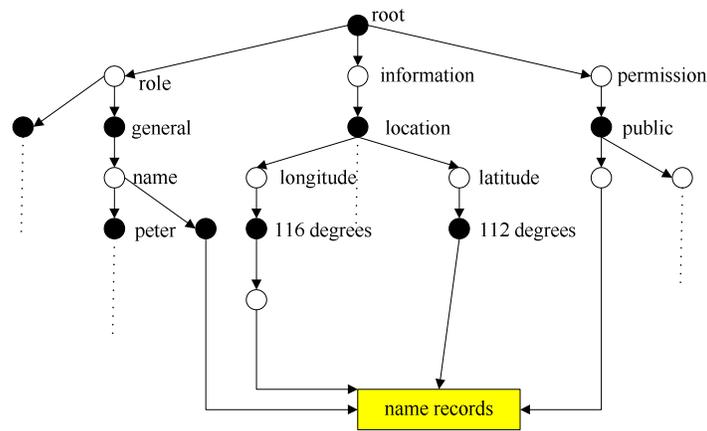

Fig.8 Name-tree storage

### 3.4 NAME Routing Algorithm

The NAME Routing Algorithm is as follows:

___________________________________________________________________

NAME Routing Algorithm
___________________________________________________________________

**Input:** the coming bundle $B_i$ ( $0 < i < m$ )

**Output:** bundles routing succeeds or fails

1. Pick a bundle $B_i$ in pending_bundle_table ( $B_i$ ):
2.    **if** dis( $B_i$.location, $B_i$.target_location) > $B_i$.target_radius:
3.       $B_i$.route_state = STEM
4.       **for** each Node $N_j$ ( $0 < j < n$ ) in node_table( $N_j$ ):
5.          **if** dis( $N_j$.location, $B_i$.target_location) < min
6.             min = dis( $N_j$.location, $B_i$.target_location)
7.             Node_tag = min_node
8.       forward $B_i$ to min_node
9.    **else**
10.       change $B_i$.route_state from STEM to FLOOD
11.       **for** each Node $N_j$ in node_table( $N_j$ ):
12.          **if** $B_i$ have a direct link to Node $N_j$
13.             forward bundle $B_i$ to Node $N_j$
14. **end**
___________________________________________________________________

The original node first uses predicate expression "within (location, distance)" to narrow the scope of target region. In other words, if the distance between the original node and a target node is greater than target_radius which is a constant value, we will choose the routing state STEM. The bundle is to be transferred to nodes that are closer to the target node. This process will not stop until the bundle locates in the target region, then the routing state changes as FLOOD. In this routing state, inside the circle defined by (x, y, distance), the bundle will be delivered to all the nodes in the list of next hop nodes, which are within the target region. In other words, it performs a lookup into its name KB and returns only those canonical EIDs that satisfy the predicate "within(x, y, distance)." The flow chart of NAME routing is displayed as follows:

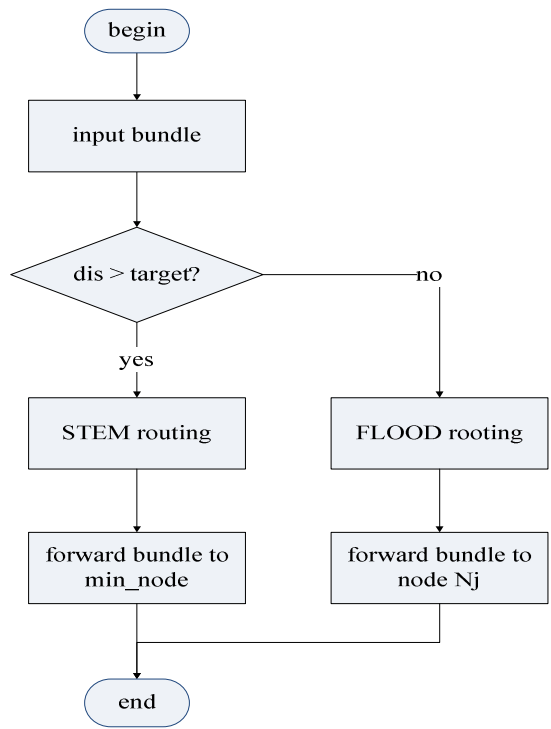

Fig.9 NAME routing flow chart

In order for name resolution schemes during NAME routing, location information needs to be shared among nodes in DTN network. Each node broadcasts location update bundle periodically to all nodes. When nodes receive location update bundles from neighbour nodes, they will update (location, canonical EID) pairs which are stored in databases.

## 4    Conclusion

In this paper, we first discuss the whole architecture of DTN naming mechanism. And then introduce our design and implementation of NAME. Our goals are to create an expressiveness, responsiveness, robustness and easy configuration naming mechanism. NAME uses a simple naming strategy based on attributes and values to finish locating resource and name resolution. We also design a predicate-based routing algorithm which is simple and efficient.

In future work, we will complete our DTN naming system, and design APIs for all kinds of applications such as content-based data dissemination and resource location. The APIs mainly

240

include three parts: Local resolution, Delivery to application and Remote resolution. As the name ontology becomes more complex, the design and implementation of DTN naming system could be much harder than now. How to design a flexible and well performed DTN naming system is still a considerable problem.

## Acknowledgment

This work is supported by the National Science Foundation of China under Grant No. 61073076, the Science and Technology on Information Transmission and Dissemination in Communication Networks Laboratory under Grant No.ITD-U12001, and Beihang University Innovation and Practice Fund for Graduate under Grant No. YCSJ-02-02. The authors would like to thank great support.